\pdfoutput=1
\documentclass[sigconf,nonacm]{acmart}
\setcopyright{none}
\usepackage[shortlabels]{enumitem}
\usepackage{stmaryrd}
\usepackage{pgfplots}
\pgfplotsset{
	compat=1.13,
}
\usepackage{forest}
\usetikzlibrary{arrows,automata}
\usepackage{xcolor}
\usetikzlibrary{arrows.meta} 
\usetikzlibrary{angles}
\usetikzlibrary{positioning}
\definecolor{drawColor}{RGB}{128 128 128}
\newcommand{\circleSize}{0.25em}
\newcommand{\angleSize}{1em}
\forestset{
	/tikz/mandatory/.style={
		circle,fill=drawColor,
		draw=drawColor,
		inner sep=\circleSize
	},
	/tikz/optional/.style={
		circle,
		fill=white,
		draw=drawColor,
		inner sep=\circleSize
	},
	featureDiagram/.style={
		for tree={
			text depth = 0,
			parent anchor = south,
			child anchor = north,
			edge = {draw=drawColor},
            l sep = 15pt,
            s sep = 10pt,
            minimum width=4.5em,
		}
	},
    featureDiagram2/.style={
		for tree={
            l sep = 0pt
		}
	},
	/tikz/abstract/.style={
		fill = blue!85!cyan!5,
		draw = drawColor
	},
	/tikz/concrete/.style={
		fill = blue!85!cyan!20,
		draw = drawColor
	},
	mandatory/.style={
		edge label={node [mandatory] {} }
	},
	optional/.style={
		edge label={node [optional] {} }
	},
	or/.style={
		tikz+={
			\path (.parent) coordinate (A) -- (!u.children) coordinate (B) -- (!ul.parent) coordinate (C) pic[fill=drawColor, angle radius=\angleSize]{angle};
		}	
	},
	/tikz/or/.style={
	},
	alternative/.style={
		tikz+={
			\path (.parent) coordinate (A) -- (!u.children) coordinate (B) -- (!ul.parent) coordinate (C) pic[draw=drawColor, angle radius=\angleSize]{angle};
		}	
	},
	/tikz/alternative/.style={
	},
	/tikz/placeholder/.style={
	},
	collapsed/.style={
		rounded corners,
		no edge,
		for tree={
			fill opacity=0,
			draw opacity=0,
			l = 0em,
		}
	},
	/tikz/hiddenNodes/.style={
		midway,
		rounded corners,
		draw=drawColor,
		fill=white,
		minimum size = 1.2em,
		minimum width = 0.8em,
		scale=0.9
	},
}

\begin{document}
\title{Mapping Cardinality-based Feature Models to Weighted Automata over Featured Multiset Semirings}
\subtitle{(Extended Version)}

\author{Robert M{\"u}ller}
\affiliation{%
	\institution{University of Siegen}
	\city{Siegen}
	\country{Germany}
}
\email{robert.mueller@uni-siegen.de}

\author{Mathis Wei{\ss}}
\affiliation{%
	\institution{University of Siegen}
	\city{Siegen}
	\country{Germany}
}
\email{mathis.weiss@uni-siegen.de}

\author{Malte Lochau}
\affiliation{%
	\institution{University of Siegen}
	\city{Siegen}
	\country{Germany}
}
\email{malte.lochau@uni-siegen.de}

\renewcommand{\shortauthors}{M{\"u}ller, Wei{\ss} and Lochau}

\begin{abstract}
Cardinality-based feature models permit to select multiple copies of the same feature, thus generalizing the notion of product configurations from subsets of Boolean features to multisets of feature instances. This increased expressiveness shapes a-priori infinite and non-convex configuration spaces, which renders established solution-space mappings based on Boolean presence conditions insufficient for cardinality-based feature models. To address this issue, we propose weighted automata over featured multiset semirings as a novel behavioral variability modeling formalism for cardinality-based feature models. The formalism uses multisets over features as a predefined semantic domain for transition weights. It permits to use any algebraic structure forming a proper semiring on multisets to aggregate the weights traversed along paths to map accepted words to multiset configurations. In particular, tropical semirings constitute a promising sub-class with a reasonable trade-off between expressiveness and computational tractability of canonical analysis problems. The formalism is strictly more expressive than featured transition systems, as it enables upper-bound multiplicity constraints depending on the length of words. We provide a tool implementation of the behavioral variability model and present preliminary experimental results showing applicability and computational feasibility of the proposed approach.
\end{abstract}

\maketitle

\noindent © \shortauthors{} 2024. This is the author's version of the work. It is posted here for your personal use. Not for redistribution. The definitive version was published in \textit{Proceedings of 28th ACM International Systems and Software Product Lines Conference (SPLC'24)}, \url{https://doi.org/10.1145/3646548.3676539}.

\section{Introduction}\label{sec:intro}

\paragraph{Background and Motivation.}

Many modern software-intensive systems must be highly configurable
to allow flexible adaptations to user requirements, technical platforms etc. on demand.
Software product-line engineering offers a comprehensive methodology
to systematically develop a family of similar, yet explicitly distinguished 
software variants being automatically derivable from a configurable software implementation~\cite{Apel2013FeatureOrientedSP}.
The variability of a configurable software
is defined by \textit{features} (i.e., customer-visible configuration options).
Most recent product-line approaches are limited to
\textit{finite Boolean} configuration spaces, shaped by a predefined 
set $F$ of configuration options (features) from the problem domain.
A configuration $C\subseteq F$ thus corresponds to a subset
of selected features.
Feature models such as feature diagrams~\cite{kang1990feature,FIDEBook2017} are 
used to specify the (sub-)space of \textit{valid} configurations supported by a product line, 
which formally corresponds to a subset 
of the overall set $2^F$ of all possible feature selections.

Features also build the conceptual glue between problem-space 
and solution-space variability, in the sense that every valid
feature selection controls the presence/absence of 
optional increments of implemented functionality~\cite{Apel2013FeatureOrientedSP}.
Initially inspired by C pre-processor directives~\cite{LiebigCPPSPL2010}, 
most recent feature-mapping techniques attach
\textit{presence conditions} (i.e., propositional formulae over features) to
variable solution-space artifacts (e.g., model elements and code fragments)~\cite{CzarneckiTemplate2005,classen2012featured,benduhn2015survey}.
Presence conditions symbolically characterize in a concise way
the subsets of feature selections for which the marked
artifacts are included in the respective software variant.
For instance, in featured transition systems (FTS), transitions
are annotated by presence conditions to specify the subset 
of configurations for which the behavior defined by this transition is relevant~\cite{classen2012featured}.
In this way, one FTS model superimposes the set of all variants
included in the valid configuration space into one behavioral model.

However, in many modern application domains like cloud computing and cyber-physical
systems, Boolean configuration options are not sufficiently expressive. 
Customers may not only decide about the presence or absence, but also about the \textit{multiplicity} (number of instances) of resources.
In cardinality-based feature models (CFMs), features are augmented by cardinality intervals $\langle l,u\rangle$, $l\leq u$, requiring a feature 
to be selected at least $l$ and at most $u$ times~\cite{czarnecki2005cardinality}.
The notion of configurations thus generalizes to 
\textit{multisets} $M:F\rightarrow \mathbb{N}_{0}$ over feature sets $F$ and
the valid configuration space of CFMs 
comprises a subset of all possible multisets over $F$~\cite{weckesser2016mind}.
For the upper-bound $u$, an a-priori unbounded 
wildcard $*$ may be used instead of a fixed value,
which gives rise to \textit{infinite} configuration spaces.
Moreover, further constructs for expressing cardinality constraints
may lead to interval gaps within \textit{non-convex} configuration spaces~\cite{weckesser2016mind}.

\paragraph{Research Challenges.}

Adapting feature-mapping concepts 
from Boolean feature models to cardinality-based
feature models is not straight-forward and, to the best of our knowledge,
no systematic approach tackling this issue has been proposed so far.
Many structural modeling formalisms offer built-in constructs for 
declaring multi-instantiable model entities~\cite{WeckesserClafer2018}
(e.g., multiplicity constraints in UML class diagrams), whereas 
typical behavioral models like transition systems do not 
comprise an analogous counterpart.
To fill this conceptual gap, we face the following research challenges:

\smallskip
\noindent\textbf{Expressiveness-Tractability Trade-offs.}
There is no obvious generalization of the notion
of presence conditions from propositional formulae
over Boolean features to multisets of feature instances.
A naive approach would rely on some generalization of
propositional logics being rich enough
to express \textit{every possible subset of all multisets} over feature set $F$.
However, this would lead to presence conditions
being far more expressive (and far more complex to 
comprehend and to analyze) than cardinality constraints of CFMs,
thus leading to a conceptual misalignment between
problem-space and solution-space variability.

\smallskip
\noindent\textbf{Automated Reasoning.}
Presence conditions over Boolean features
are usually added to existing 
modeling/programming languages as \textit{syntactic} mark-ups~\cite{CzarneckiTemplate2005} 
or by \textit{semantic} variability encoding~\cite{PostVariabilityEncoding2008}.
Hence, their proper handling either requires a separate mechanism complementing
native tool support of the host language or a lifting of
existing tools to properly handle variability.
Automated reasoning about Boolean presence conditions 
is usually delegated to constraint solvers~\cite{BatoryFM2SAT2005,MendoncaFMSATeasy2009}. 
For multiset-based presence conditions, however, 
no such canonical mathematical representation 
with out-of-the-box analysis tools exists.    

To define a feasible mapping approach
between CFMs and behavioral solution-space
modeling, we consider the following research challenges.

\begin{itemize}
    \item Define a mapping approach that constitutes a reasonable trade-off between expressiveness and analyzability.
    \item Define a mapping approach that enables a semantic embedding
    of feature multiplicities into an existing behavioral host formalism with native tool support.
\end{itemize}

\paragraph{Concepts and Contributions.}
In this paper, we argue that \textit{weighted automata}~\cite{droste2009handbook} 
constitute a suitable behavioral modeling formalism for
mapping configuration spaces of CFMs to
behavioral variability in the solution space.
Weighted automata generalize classical finite automata
by assigning weights to accepted words, where
the domain of weights applied to a 
weighted automaton can be any algebraic structure 
constituting a \textit{semiring}~\cite{manger2008catalogue}.
Here, we employ weighted automata over \textit{multisets} as target
behavioral model for solution-space mapping of CFMs.
Transitions are annotated by multisets $M$ over
feature set $F$, denoting the \textit{multiplicity} of feature instances 
\textit{at least} required for this transition to be enabled.
In this way, we obtain a natural semantic mapping between
multisets of features corresponding to valid configurations of a CFM
in the problem space and multisets of features assigned 
to behaviors (words) of a weighted automaton in the solution space.
This conceptual correspondence allows us
(1) to apply arbitrary semiring-preserving operations on multisets
to aggregate feature multiplicity constraints in different ways, and
(2) to adopt many useful analysis techniques known for 
certain classes of weighted automata~\cite{almagor2022s}
to efficiently reason about key properties in a family-based and 
an automated way~\cite{ThuemSPLAnalysisSurvey2014}.
In particular, using tropical semirings over multisets appears to
be a reasonable trade-off between expressiveness and complexity. 

\smallskip
\noindent\textbf{Tool Support.}
We provide tool support based on the JAutomata library~\cite{JAutomata}
which allows the creation of behavioral model instances and provides the automatic analysis using the proposed formalism.

\smallskip
\noindent\textbf{Experimental Evaluation.}
We describe experimental results obtained by
applying our tool to a collection of example
models demonstrating efficiency/effectiveness trade-offs.

\smallskip
\noindent\textbf{Verifiability.}
To make our results reproducible, we provide the
tool implementation, all experimental results and raw
data on a supplementary web page~\cite{mueller2024mappingartifact}.
\section{Background: Cardinality-Based Feature Models}\label{sec:background}

We introduce a running example to illustrate the modeling concepts of cardinality-based feature models.

\smallskip
\noindent\textbf{Running Example: Multiplayer Game.}
We consider a simplified multiplayer game
inspired by~\cite{richerzhagen2014bypassing, sampling2024}.
We do not describe the actual gameplay in detail,
but focus on the configuration of the game by players.
In each game, two, up-to arbitrary many, teams 
participate in a game of solitaire or chess.
Each team has at least one player, whereas 
the maximum number of players per team is not limited.
Processing modules provide configurable
communication protocols for each team (e.g., WiFi or Bluetooth).
Some configuration constraints can be specified in a 
classical (Boolean) feature model 
(e.g., alternative choice between solitaire and chess)~\cite{kang1990feature}. 
In contrast, the fact that the number of teams is
freely configurable and that each such team can have 
an individually configurable number of players is not expressible in classical 
Boolean feature models.
Instead, we employ cardinality-based feature models (CFMs) to precisely specify
the valid configuration space (problem space) of the game.

\smallskip
\noindent\textbf{Syntax of CFM.}
In classical (Boolean) feature models, each feature 
is either selected (true) or deselected (false) 
in a configuration~\cite{kang1990feature} and the
valid configuration space is shaped by propositional
constraints over features. CFMs generalize this 
by feature \textit{multiplicities} to denote the number of times a feature 
is selected~\cite{czarnecki2005formalizing, czarnecki2005cardinality}.
The valid configuration space of CFMs is thus shaped by cardinality constraints
restricting valid multiplicity intervals of feature instances.

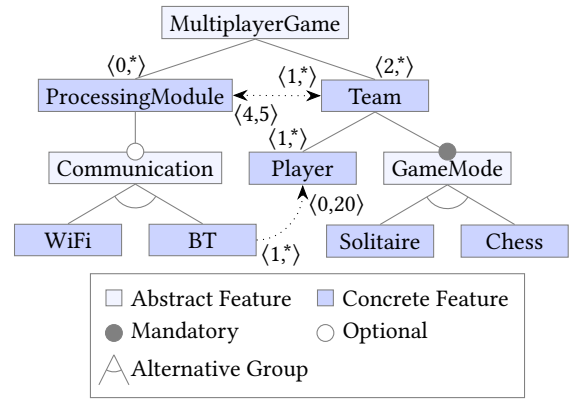
\begin{figure}[pt]
    \centering
    \begin{forest}
	featureDiagram
	[MultiplayerGame, abstract
		[ProcessingModule, name=procMod, concrete, edge label={node[very near end, left, xshift=2pt, yshift=3pt]{$\langle$0,*$\rangle$}}
			[Communication, abstract, optional
				[WiFi, concrete, alternative]
				[BT, name=bt, concrete]
			]
		]
		[Team, name=team, concrete, edge label={node[very near end, right, xshift=2pt, yshift=3pt]{$\langle$2,*$\rangle$}}
			[Player, name=player, concrete, edge label={node[very near end, left, xshift=2pt, yshift=3pt]{$\langle$1,*$\rangle$}}]
			[GameMode, abstract, mandatory
				[Solitaire, concrete, alternative]
				[Chess, concrete]
			]
		]
	]
	\draw[-{Stealth[length=2mm]}, dotted] (bt) to[out=east,in=south] node[near start, below, xshift=2pt]{$\langle$1,*$\rangle$} node[near end,right]{$\langle$0,20$\rangle$} (player);
	\draw[{Stealth[length=2mm]}-{Stealth[length=2mm]}, dotted] (team) to[out=west,in=east] node[near start, above]{$\langle$1,*$\rangle$} node[near end,below]{$\langle$4,5$\rangle$} (procMod);
	\matrix[anchor=north west] at (current bounding box.north east) {
		\node[placeholder] {};\\
	};
	\matrix[draw=drawColor,anchor=north] at (current bounding box.south) {
		\node[abstract,label=right:Abstract Feature] {}; \pgfmatrixnextcell
		\node[concrete,label=right:Concrete Feature] {};\\
		\node[mandatory,label=right:Mandatory] {}; \pgfmatrixnextcell
		\node[optional,label=right:Optional] {};\\
			\draw[drawColor] (0,0.1) -- +(-0.2, -0.4);
			\draw[drawColor] (0,0.1) -- +(0.2,-0.4);
			\draw[drawColor] (-0.1,-0.1) arc (240:300:0.2);
		\node[alternative,label=right:Alternative Group] {};\\	};
\end{forest}%
    \caption{Cardinality-Based Feature Model for the Multiplayer Game.}
    \label{fig:cfm-example-mgs}
\end{figure}

Figure~\ref{fig:cfm-example-mgs} shows a CFM for our running example.
The cardinality constraint $\langle2,*\rangle$ attached at feature \textit{Team} restricts the multiplicity interval for it.
The lower-bound value $2$ denotes that each valid multiplayer game configuration must consist of at least two teams 
(i.e., two instances/copies/clones of feature \textit{Team}). 
As upper-bound value, the wildcard $*$ denotes
that an arbitrary (a-priori unbounded) number of instances 
of feature \textit{Team} may be chosen.
For each instance, the corresponding
sub-tree of feature \textit{Team} is cloned to enable individual
configuration decisions (e.g., number of players) for each team instance.
Sub-feature \textit{Player} is annotated with
$\langle1,*\rangle$, thus allowing at least one, up-to an arbitrary number of players per team.

We use slightly simplified CFM syntax to keep the example graspable.
For instance, feature \textit{GameMode} is marked as a mandatory 
sub-feature of feature \textit{Team} using
FODA notation as a shorthand for CFM constraint $\langle1,1\rangle$.
For each team, exactly one game mode must 
be selected from the alternative group (in FODA notation) containing 
\textit{Solitaire} and \textit{Chess}.
CFM generalizes the notion of groups by two kinds of group cardinality constraints~\cite{weckesser2016mind} which we also omit here for the sake of simplicity.
Also, cross-tree constraints like require- and exclude-edges are
generalized in CFM~\cite{quinton2013cardinality}. 
In our example, an exclude-edge holds between features \textit{Processing Module} and \textit{Team}.
The cardinality constraints attached to both ends of the edge
express that if $1$ to $*$ many teams are selected, 
then we cannot select $4$ or $5$ processing modules and vice versa.
We added this made-up constraint for illustrative purposes only.
The require-edge from \textit{BT} to \textit{Player} expresses
that the usage of Bluetooth is only valid for teams with up to $20$ players. 

\smallskip
\noindent\textbf{Semantics of CFM.}
In case of a Boolean feature model \textit{bm} over a set $F$ 
of Boolean features, a \textit{configuration} is 
a subset $C\subseteq F$ of selected features such that all constraints 
of \textit{bm} are satisfied.
Hence, the semantics of \textit{bm} classifies
a particular subset $\llbracket \textit{bm}\rrbracket\subseteq 2^F$ as 
\textit{valid configuration space}.
Abstract features (e.g., root feature \textit{MultiplayerGame} in Figure~\ref{fig:cfm-example-mgs}) 
do not represent actual functionality but are used to structure the feature model.
Thus, abstract features may be omitted in configurations $C$ as 
there is no solution-space mapping defined for them.
In CFMs, a configuration not only 
comprises information about presence/absence of features, but also
about the multiplicity (number of instances) of each feature.
The semantic domain is thus generalized from subsets to 
\textit{multisets} over $F$, denoted by a mapping $M:F\rightarrow\mathbb{N}_{0}$ 
of features to non-negative integers.
By $M(f)$, we denote \textit{multiplicities} of $f\in F$ in $M$.
We write
$$M = \{ f_{1}^{M(f_1)},\,f_{2}^{M(f_2)},\,...,\,f_{n}^ {M(f_n)} \}$$
to explicitly enumerate multisets $M$ over $F$ in a set-oriented way and omit entries for those features $f_i\in F$ with $M(f_i) = 0$.
A valid configuration with two teams having 
each one player playing chess together without
processing modules corresponds to the multiset:
\[M=\{Team^2, Player^2, Chess^2\}\text{,}\]
omitting abstract features.
In contrast, the multiset:
\[M'=\{BT^1, ProcessingModule^4, Team^2, Player^{30}, Solitaire^1\}\]
does not correspond to a valid configuration 
as it violates various constraints of the CFM
(e.g., the exclude-edge is violated as $4$
instances of \textit{ProcessingModule} are selected).
The valid configuration space shaped by CFMs such as our example has two remarkable properties as compared to Boolean feature models.
\begin{itemize}
    \item \textbf{Infinite.}
    The number of players, teams and processing modules are a-priori unbounded (upper bound $*$) thus the valid configuration space contains infinitely many configurations.
    \item \textbf{Non-convex.}
    Due to the exclude-edge between \textit{Team} 
    and \textit{ProcessingModule}, the sub-interval $[4..5]$ is a gap within the valid interval. 
    The resulting non-convexity of the valid configuration space makes automated analysis harder than in the convex case~\cite{weckesser2016mind}.
\end{itemize}
CFMs comprise further semantic subtleties for which no generally accepted interpretation exists.
For instance, we consider the example configuration $M$ described
above as valid, as we assume that both teams \textit{may} consist of exactly one player.
However, it may also be possible that in $M$ both players are part
of the \textit{same} team instance which would be invalid.
The first interpretation (\textit{global} multiset-based) only 
comprises the total number of instances per feature, whereas
the second interpretation (\textit{local} instance-based)
comprises a richer notion of configuration
including instances of child-parent relationships on feature instances.
For the sake of simplicity, we consider the global interpretation
in this paper (see~\cite{sampling2024} for more details).
In this case, the semantics of a CFM $\textit{cm}$
classifies a subset $\llbracket \textit{cm}\rrbracket\subseteq 2^{\mathcal{M}_{F}}$
of the set $\mathcal{M}_{F}=\{M\,|\,M:F\rightarrow \mathbb{N}_{0}\}$
of all multisets over feature set $F$ as valid.
However, this semantic generalization from Boolean feature models
to cardinality-based feature models is not straight-forward as 
the semantic domain of $\mathcal{M}_{F}$ essentially comprises \textit{all possible}
mappings from feature sets $F$ into the non-negative integers.
In this sense, the expressiveness of CFM is far away from being complete. 
For example, we cannot model a game in which we allow 
an arbitrary, but \textit{even} number of players in each team.
These and more fine-grained configuration constraints
must be expressed by behavioral variability models in the solution space instead.
\section{Mapping Cardinality-Based Feature Models to FTS}

We now discuss the limitations of a state-of-the-art behavioral variability modeling approach for finite, Boolean configuration spaces to serve as a mapping target for configuration spaces shaped by cardinality-based feature models.

\smallskip
\noindent\textbf{State-of-the-Art: Featured Transition Systems.}
\begin{figure}[pt]
    \centering
    \begin{tikzpicture}[->, >={Stealth[length=2mm]}, auto, semithick]
    \tikzstyle{every state}=[fill=none, draw=black, text=black]

    \node[state, initial]     (A)                       {$q_1$};
    \node[state]              (B) [right=2.5cm of A]    {$q_2$};
    \node[state]              (C) [below=1.3cm of B]    {$q_3$};
    \node[state]              (D) [left=2.5cm of C]     {$q_4$};
    \node[state, accepting]   (E) [left=2.2cm of D]     {$q_5$};

    \path (A) edge              node[above, align=left, yshift=-2pt]                {addTeam\\$\lbrack Team\rbrack$}                            (B)
          (B) edge [loop right] node[above, align=left, xshift=5pt, yshift=1pt]     {addPlayer\\$\lbrack Player\rbrack$}                        (B)
              edge              node[right, align=left]                             {addTeam\\$\lbrack Team\rbrack$}                            (C)
              edge [bend left]  node[left, align=right, xshift=13pt, yshift=17pt]   {addSolitaire\\$\lbrack Player \land Solitaire\rbrack$}     (D)
          (C) edge [loop right] node[below, align=left, xshift=5pt, yshift=-3pt]    {addPlayer\\$\lbrack Player\rbrack$}                        (C)
              edge              node[below, align=center, yshift=2pt]               {addChess\\$\lbrack Player \land Chess\rbrack$}             (D)
          (D) edge [bend left]  node[below, align=center, yshift=2pt]               {addBT\\$\lbrack ProcMod \land BT\rbrack$}                  (E)
              edge              node[above, align=center, yshift=-2pt]              {addWiFi\\$\lbrack WiFi\rbrack$}                            (E)
              edge [bend left]  node[left, align=right, yshift=7pt]                 {addProcMod\\$\lbrack ProcMod\rbrack$}                      (A);
\end{tikzpicture}%
    \caption{(F)TS for the Multiplayer Game.}
    \label{fig:fts-example}
\end{figure}
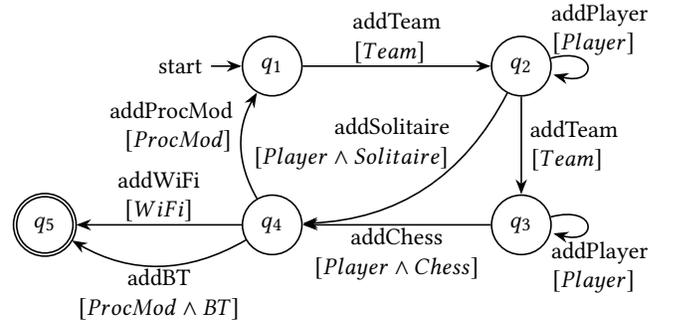
A finite state automaton $\textit{A}$ defined over an alphabet
$\Sigma$ characterizes a \textit{language} $\mathcal{L}($\textit{A}$)\subseteq \Sigma^{*}$ containing only those words $\ell\in\Sigma^{*}$ (i.e., finite sequences of symbols from $\Sigma$) for which an accepting path exists in $\textit{A}$.
In this sense, $\textit{A}$ defines a function $\textit{A}:\Sigma^{*}\rightarrow \{\bot,\top\}$ (i.e., from words to Booleans).
Transition systems formalize the possible behavior (sets of runs)
of computational systems as sequences of steps between
states via transitions, labeled by actions.
Figure~\ref{fig:fts-example} shows a transition system of our example.
All runs start in the initial state $q_{1}$
and terminate in the final state $q_{5}$.
Initially, one team is added to the game, followed by an arbitrary
number of players.
Then, the solitaire game mode can be selected,
which requires at least one team.
Alternatively, another team with additional players can be added.
In this branch, also selecting the chess game mode becomes possible,
as this requires two teams.
After one game has been completely configured, an additional processing module
may be added to the system to configure more games.
Finally, either Bluetooth or WiFi must be selected.

Transition systems have been extended to
Featured Transition Systems (FTS) to map
problem-space variability 
to behavioral variability in the solution space~\cite{classen2012featured}.
FTS extend transition labels by a second
component denoting \textit{presence conditions}.
A presence condition is a propositional formula over
Boolean features that symbolically specifies for the corresponding transition
the sets of configurations for which this transition is present.
In Figure~\ref{fig:fts-example}, presence
conditions are denoted in square brackets.
In this sense, FTS are the behavioral complement of Boolean configuration spaces
due to the conceptual alignment between Boolean feature models and
Boolean presence conditions over features.
For instance, the choice between Bluetooth or WiFi communication 
maps directly to the alternative branches in the FTS 
with mutually excluding presence conditions.

Semantically, the logical conjunction of all presence conditions 
along the path taken in a run (accepted word) qualifies the configurations for which this run is \textit{valid}.
FTS therefore define a mapping from the set of runs 
to a subset of valid configurations.
For instance, the word \textit{addTeam, addSolitaire, addWiFi}
is mapped to path condition $Team \land Player \land Solitaire \land WiFi$
thus requiring at least these features to be selected.

\smallskip
\noindent\textbf{Expressing Multiplicity Constraints in FTS.}
FTS with Boolean presence conditions are able to express that the absence (instance multiplicity $0$) or presence (instance multiplicity $>0$) of particular \textit{types} of features is required for a particular transition.
In contrast, multiplicity constraints denoting that an exact and/or
relative number of feature instances is required for a transition
are not directly expressible in Boolean FTS.
For instance, the Boolean presence condition of the transition labeled 
\textit{addBT} specifies that Bluetooth 
requires \textit{at least one} processing module. 
However, this does not reflect that \textit{sufficiently many} processing modules must be selected within the \textit{same} run.
We next discuss two possible solutions
within the FTS framework: (1) duplication of the state-transition graph
and (2) extension of presence conditions to higher-order logics.

\begin{figure}[pt]
    \centering
      \begin{tikzpicture}[->, >={Stealth[length=2mm]}, auto, semithick, align=center, node distance=1.5cm]
    \tikzstyle{every state}=[fill=none, draw=black, text=black]

    \node[state, initial]     (A)                         {$q_1^1$};
    \node[state]              (B) [right=2cm of A]        {$q_2^1$};
    \node[state]              (C) [below of=B]            {$q_3^1$};
    \node[state]              (D) [left=2cm of C]         {$q_4^1$};
    \node[state, accepting]   (E) [left=2cm of D]         {$q_5^1$};

    \node[state]              (A1) [below=1cm of D]       {$q_1^2$};
    \node[state]              (B1) [right=2cm of A1]      {$q_2^2$};
    \node[state]              (C1) [below of=B1]          {$q_3^2$};
    \node[state]              (D1) [left=2cm of C1]       {$q_4^2$};
    \node[state, accepting]   (E1) [left=2cm of D1]       {$q_5^2$};

    \node[state]              (A2) [below=1cm of D1]      {$q_1^3$};
    \node[state]              (B2) [right=2cm of A2]      {$q_2^3$};
    \node[state]              (C2) [below of=B2]          {$q_3^3$};
    \node[state]              (D2) [left=2cm of C2]       {$q_4^3$};
    \node[state, accepting]   (E2) [left=2cm of D2]       {$q_5^3$};

    \path (A)     edge                node[above, align=left]                             {addTeam}                                   (B)
          (B)     edge [loop right]   node[above, align=left, xshift=4pt, yshift=1pt]     {addPlayer}                                 (B)
                  edge                node[right, align=left]                             {addTeam}                                   (C)
                  edge                node[left, align=right, xshift=-5pt]                {addSolitaire}                              (D)
          (C)     edge [loop right]   node[below, align=left, xshift=4pt, yshift=-4pt]    {addPlayer}                                 (C)
                  edge                node[below, align=center]                           {addChess}                                  (D)
          (D)     edge [bend left]    node[below, align=center]                           {addBT\\$\lbrack ProcMod \rbrack$}          (E)
                  edge                node[right, align=left]                             {addProcMod\\$\lbrack ProcMod \rbrack$}     (A1)
          (A1)    edge                node[above, align=left]                             {addTeam}                                   (B1)
          (B1)    edge [loop right]   node[above, align=left, xshift=4pt, yshift=1pt]     {addPlayer}                                 (B1)
                  edge                node[right, align=left]                             {addTeam}                                   (C1)
                  edge                node[left, align=right, xshift=-5pt]                {addSolitaire}                              (D1)
          (C1)    edge [loop right]   node[below, align=left, xshift=4pt, yshift=-4pt]    {addPlayer}                                 (C1)
                  edge                node[below, align=center]                           {addChess}                                  (D1)
          (D1)    edge [bend left]    node[below, align=center]                           {addBT\\$\lbrack ProcMod \rbrack$}          (E1)
                  edge                node[above, align=center]                           {addWiFi}                                   (E1)
                  edge                node[right, align=left]                             {addProcMod\\$\lbrack ProcMod \rbrack$}     (A2)
          (A2)    edge                node[above, align=left]                             {addTeam}                                   (B2)
          (B2)    edge [loop right]   node[above, align=left, xshift=4pt, yshift=1pt]     {addPlayer}                                 (B2)
                  edge                node[right, align=left]                             {addTeam}                                   (C2)
                  edge                node[left, align=right, xshift=-5pt]                {addSolitaire}                              (D2)
          (C2)    edge [loop right]   node[below, align=left, xshift=4pt, yshift=-4pt]    {addPlayer}                                 (C2)
                  edge                node[below, align=center]                           {addChess}                                  (D2)
          (D2)    edge                node[above, align=center]                           {addWiFi}                                   (E2);
\end{tikzpicture}%
    \caption{FTS Representation with Duplicated Parts.}
    \label{fig:fts-example-unrolled}
\end{figure}
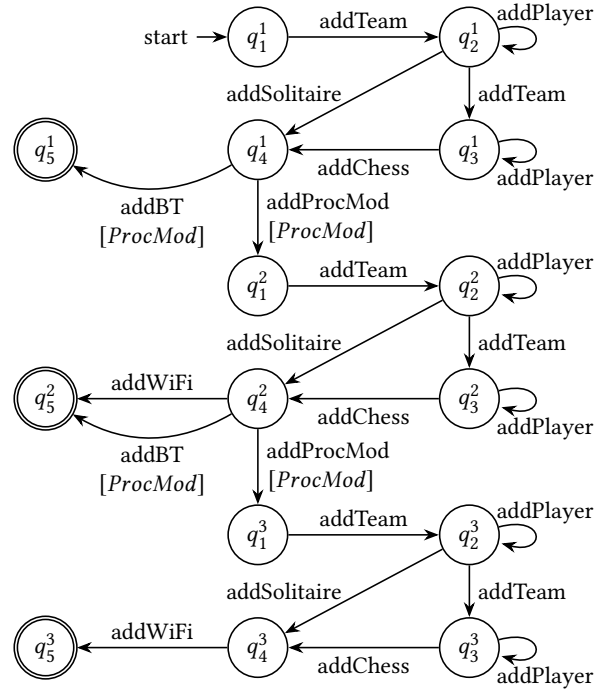

\noindent\textit{(1) Duplicating FTS Parts.}
The CFM in Figure~\ref{fig:cfm-example-mgs}
allows selecting any number $k$ of processing modules
in a valid multiset configuration $M$.
For instance, in case of $k=2$,
only those runs of the FTS should be considered
valid for $M$ in which the action \textit{addProcMod} occurs
at most two times which is not expressible by Boolean presence conditions.
Figure~\ref{fig:fts-example-unrolled}
shows a possible FTS representation for the case $k=2$ using
duplicated parts.
Note that all presence conditions except for those involving \textit{ProcMod}
have been omitted here for brevity.
The transition loop for \textit{addProcMod} has been unrolled 
twice to ensure that the \textit{addProcMod} transition is taken 
at most two times in any valid run.
After taking the \textit{addProcMod} transition twice,
choosing \textit{Bluetooth} becomes disabled, as Bluetooth
requires an additional processing unit.
Additionally, the transition for \textit{addWiFi} from state 
$q_4^1$ to state $q_5^1$ has been removed
as the \textit{WiFi} feature requires at least one processing module,
which does not hold after the first iteration.
The last transition for \textit{addBT} from state $q_4^3$ to state $q_5^3$ has also been removed as this would require a further processing module 
which is not permitted by configuration $M$.
This pattern may be applied for any given upper-bound value $k$, 
yet leading to excessively bloated behavioral models.
Even worse, due to the explicitly declared unboundedness of 
feature \textit{ProcMod} in the CFM, no such upper-bound value exists,
thus yielding an infinite number of unfoldings which is
infeasible as (F)TS are limited to a finite number of states.

\smallskip
\noindent\textit{(2) Extending Presence Conditions.}
Another approach consists in a generalization of 
presence conditions from propositional formulas 
over Boolean formulas to higher-order 
logical formulas including multisets as modulo theories.
For instance, the extended presence condition
$[\textit{ProcMod}^{2} \wedge \textit{WiFi}^{1}]$
attached to a transition for \textit{addWiFi}
would require one instance of feature \textit{WiFi} and
two instances of feature \textit{ProcMod}.
However, regardless of the expressiveness of the formalism used for 
presence conditions to mark \textit{individual transitions},
the state space of FTS remains limited to a finite number of states.
In this regard, the notion of presence conditions 
used in FTS-like behavioral variability models 
only enables \textit{qualitative} mapping constraints 
for individual transitions.
More involved mapping constraints including aggregated conditions over 
\textit{complete runs} are not expressible in this way.
For instance, we cannot express
that the number of times the \textit{addProcMod} is taken in a run
requires a configuration with exactly this number of \textit{ProcMod} feature instances.
Furthermore, FTS-like presence conditions cannot express \textit{quantitative} constraints
like aggregating properties related to feature multiplicities reflecting
costs or rewards caused by feature selections.
Enabling FTS-like presence conditions to also express these types of constraints
and properties would require far more complicated extensions,
such as quantifiable multiplicity variables and (hyper-)properties
over aggregated multiplicities.
If these extensions are not introduced with care, 
we obtain an incomprehensible and computationally intractable mapping formalism
which obstructs usability and automated analysis.

\smallskip
\noindent\textbf{Summary.} A straight-forward generalization of presence conditions of FTS-like behavioral models to handle configuration spaces of CFM is not obvious for two reasons.
First, configuration semantics of Boolean feature models denotes purely \textit{qualitative} constraints, being conceptually aligned with Boolean presence conditions (propositional formulae over features) in the solution space.
In contrast, configuration semantics of cardinality-based
feature models denotes \textit{quantitative} constraints (multiplicity of feature instances).
Second, FTS-like presence conditions are
usually added as conservative (i.e., purely \textit{syntactic}) extensions to existing solution-space modeling/programming languages, being (post-)processed by separate tools (e.g., SAT solvers).
In contrast, a proper mapping of feature instances asks 
for a conceptually integrated (i.e., \textit{semantic}) embedding
into an appropriate solutions-space formalism.
\section{Mapping Cardinality-Based Feature Models to Weighted Automata}\label{sec:main}

In this section, we describe a solution-space mapping for cardinality-based
feature models using weighted automata over multisets.

\subsection{Background: Weighted Automata}

Similar to FTS, weighted automata also constitute a conservative extension
of finite state automata~\cite{droste2009handbook}.
A weighted automaton $\mathcal{A}:\Sigma^{*}\rightarrow K$ generalizes finite state automata by mapping words $\ell\in \Sigma^{*}$ to arbitrary sets $K$ denoting \textit{weights}~\cite{droste2021weighted, droste2009handbook, almagor2022s}. 
Set $K$ is part of a \textit{semiring} comprising two binary operations, $\oplus$ (\textit{addition}) and $\otimes$ (\textit{multiplication}), on $K$.
The addition operation is associative and commutative
with an identity element $\overline{0} \in K$ and
the multiplication operation is associative
with an identity element $\overline{1} \in K$.
\begin{definition}[Semiring~\cite{droste2009handbook}]\label{def:semiring}
A \textit{semiring} is tuple $\mathbb{K} = (K, \oplus, \otimes, \overline{0}, \overline{1})$, where
\begin{itemize}
    \item $K$ is a non-empty set,
    \item $\oplus : K \times K \rightarrow K$
    s.t. $(a \oplus b) \oplus c = a \oplus (b \oplus c)\,\,\,$
    $a \oplus b = b \oplus a$,
    \item $\otimes : K \times K \rightarrow K$
    such that $(a \otimes b) \otimes c = a \otimes (b \oplus c)$,
    \item $\overline{0}\in K$ such that $\overline{0} \oplus a = a \oplus \overline{0} = a$ and $\overline{0} \otimes a = a \otimes \overline{0} = \overline{0}$, and
    \item $\overline{1} \in K$ such that $\overline{1}\otimes a = a \otimes \overline{1} = a$.
\end{itemize}
\end{definition}
Transitions of weighted automata are labeled by pairs of symbols $\sigma\in\Sigma$ and weights $k\in K$ and states are labeled by pairs of initial and final weights, respectively.
\begin{definition}[Weighted Automaton~\cite{droste2009handbook}]\label{def:wa}
A \textit{weighted automaton} is a tuple 
$\mathcal{A} = (\mathbb{K}, \Sigma, Q, w_{i}, w_{f}, w_{t})$, where
\begin{itemize}
    \item $\mathbb{K}$ is a semiring over a set $K$,
    \item $\Sigma$ is a finite alphabet,
    \item $Q$ is a non-empty, finite set of states,
    \item $w_{i} : Q \rightarrow K$ assigns initial weights to states,
    \item $w_{f} : Q \rightarrow K$ assigns final weights to states, and
    \item $w_{t} : Q \times \Sigma \times Q \rightarrow K$
    assigns weights to transitions.
\end{itemize}
\end{definition}
The operations $\otimes$ and $\oplus$ 
compute for any word $\ell\in\Sigma^*$ its weight:
for every path labeled $\ell$, the initial weight of the first state, the final weight of the last state and all weights along the transitions of all paths labeled $\ell$ are aggregated by $\otimes$. 
The overall weight of $\ell$ is aggregated by applying $\oplus$ to the aggregated weights of each such path.
The set of all weighted words defines the \textit{weighted
language} of the automaton.

\begin{figure}[pt]
    \centering
    \begin{tikzpicture}[->, >={Stealth[length=2mm]}, auto, node distance=3.5cm, semithick, align=center]
    \tikzstyle{every state}=[fill=none]

    \coordinate[]                     (Ai);
    \node[state]                      (A) [right=0.5cm of Ai]   {$q_1$};
    \coordinate[below=0.5cm of A]     (Af);
    \node[state]                      (B) [right=1cm of A]      {$q_2$};
    \coordinate[below=0.5cm of B]     (Bf);
    \coordinate[right=0.5cm of B]     (Bi);

    \path (Ai)   edge                node[above, align=center]     {$2$}         (A)
          (A)    edge [loop above]   node[above, align=center]     {$a, 2$}      (A)
          (A)    edge                node[left]                    {$0$}         (Af)
          (A)    edge                node[above, align=center]     {$a, 1$}      (B)
          (B)    edge [loop above]   node[above, align=center]     {$a, 1$}      (B)
          (B)    edge                node[left]                    {$5$}         (Bf)
          (Bi)   edge                node[above, align=center]     {$1$}         (B);
\end{tikzpicture}%
    \caption{Example of a Weighted Automaton.}
    \label{fig:wa-simple-example}
\end{figure}
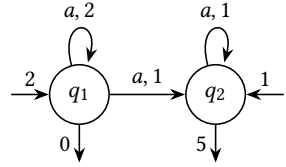

\begin{definition}[Weighted Language~\cite{droste2009handbook}]\label{def:wlang}
Let $\mathcal{A}$ be a weighted automaton over
alphabet $\Sigma$ and semiring $\mathbb{K}$.
\begin{itemize}
    \item A \textit{path} of $\mathcal{A}$ is any sequence $P = q_{0} a_{1} q_{1} ... a_{n} q_{n} \in Q(\Sigma Q)^{*}$.
    \item The \textit{weight} of path $P$ is 
    $$w(P) = w_{i}(q_{0})
    \cdot \left(\prod_{0 \leq i < n} w_{t}(q_{i}, a_{i+1}, q_{i+1})\right)
    \cdot w_{f}(q_{n})$$ 
    where the empty product is defined as $\overline{1}$.
    \item The label of a path $P$ is defined as $\ell(P) = a_{1} a_{2} ... a_{n}$.
    \item The weighted language of $\mathcal{A}$ is defined as 
    $$\mathcal{L}(\mathcal{A}) = \{ (\ell,w(\ell))\,|\, \ell \in \Sigma^{*} \wedge w(\ell) = \sum_{P\ \text{s.t.}\ \ell(P) = \ell} w(P)\}\text{.}$$
\end{itemize}
\end{definition}
\noindent\textbf{Example.} Figure~\ref{fig:wa-simple-example} shows a weighted automaton $\mathcal{A}$ defined over 
alphabet $\Sigma = \{a\}$ and semiring $\mathbb{K} = (\mathbb{N}_{0}\cup\{-\infty\}, \text{min}, +, -\infty, 0)$
using a common graphical representation.
Transitions between states are labeled by pairs $\sigma, w$ of
symbols $\sigma$ from alphabet $\Sigma$ (i.e., $\sigma=a$ in this example) and transition weights $w\in K$ defined by $w_{t}$.
The example contains three transitions: 
one from state $q_1$ to state $q_2$ with weight $1$ and two self-loop transition, one of state $q_1$ with weight $2$ and one of state $q_2$ with weight $1$.
Each state $q$ additionally has one dangling incoming edge
labeled with the initial weight
$w_{i}(q)$ and a dangling outgoing 
edge labeled with the final weight $w_{f}(q)$.
In the considered semiring, the $\otimes$-operation is defined as numerical addition of the weights (natural numbers) along paths. 
Any path for a word $\ell \in \{a\}^{*}$ 
of length $|\ell| = k$ may start
either in state $q_{1}$ or $q_{2}$.
When starting in $q_{1}$, the path 
may switch from state $q_{1}$ to state $q_{2}$
after any prefix of $\ell$ of arbitrary length $k'\leq k$ and then
remain in $q_{2}$ or it may remain in $q_{1}$ for the whole word.
The initial weight and the weight added for every additional $a$ is $2$
in $q_{1}$ and $1$ in $q_{2}$, whereas
the final weight is $0$ in $q_0$ but $5$ in $q_{2}$.
Hence, we have multiple possible paths with different aggregated weights for the same word $\ell \in \{a\}^{*}$ in $\mathcal{A}$, where the $\oplus$-operation is defined to select the \text{minimum} weight.
The weighted language of $\mathcal{A}$ for the considered semiring is
\begin{equation*}
\mathcal{L}(\mathcal{A}) = \{ (\ell, 2k + 2)\,|\,
\ell = a^{k} \wedge\,k < 5\} \cup 
\{ (\ell', k' + 6)\,|\,
\ell' = a^{k'} \wedge\,k'\geq 5\}
\end{equation*}
In all upcoming examples, we use a simplified graphical notation as follows. 
If $w_{i}(q)=\overline{1}$ holds for exactly one
state $q$ and $w_{i}(q)=\overline{0}$ for all other states $q'$
we then omit the weight
on the dangling incoming edge of \textit{initial state} $q$ (e.g., $q_1$ in Figure\ref{fig:wa-example}) 
and we also omit the dangling incoming edges on all other states $q'$.
Similarly, we omit outgoing dangling edges 
for all states $q$ with $w_{f}(q)=\overline{1}$ and instead depict $q$
as \textit{accepting state} marked by a double-lined circles (e.g., $q_5$ in Figure~\ref{fig:wa-example}), whereas all 
other states $q'$ with $w_{f}(q')=\overline{0}$ are drawn with single-lined circles.

Note that any finite state automaton is a weighted automaton over 
the Boolean semiring $\mathcal{B} = (\{\bot, \top\}, \lor, \land, \bot, \top)$.
Due to the generic definition of weighted automata
over any kind of semiring, it enjoys a wide range of different applications
(e.g., reasoning about non-functional properties,
natural language processing, image processing)~\cite{droste2009handbook}.
In particular, the \textit{tropical} semiring has proven very useful for many 
domains, where the \textit{min}-tropical semiring is defined as 
$\mathcal{T}_{min} = (\mathbb{R}\cup\{\infty\}, \text{min}, +, \infty, 0)$ 
and the \textit{max}-tropical semiring
is defined as $\mathcal{T}_{max} = (\mathbb{R}\cup\{-\infty\}, \text{max}, +, -\infty, 0)$, respectively \cite{pin1998tropical, almagor2022s}.
Using tropical semirings, weights along paths are added and the final weight for a word either corresponds to the minimum or maximum aggregated path weight for that word.

\subsection{Weighted Automata over Featured Multisets}

In Section~\ref{sec:background}, we described the semantic
domain of CFMs models as sets $\mathcal{M}_{F}$ of multisets over feature set $F$ (featured multisets).
Hence, it is reasonable to investigate weighted automata
over featured multisets as behavioral solution-space formalism
for CFMs.
Operations like union, intersection, and addition are 
lifted to multisets as component-wise maximum, minimum, 
and addition on the multiplicity domain $\mathbb{N}_{0}$:
\begin{itemize}
  \item $(M_1 \cup M_2)(a) = \text{max}(M_1(a), M_2(a))
    \quad\forall a \in A$,
    \item $(M_1 \cap M_2)(a) = \text{min}(M_1(a), M_2(a))
    \quad\forall a \in A$, and
    \item $(M_1 + M_2)(a) = M_1(a) + M_2(a)
    \quad\forall a \in A$,
\end{itemize}
To generalize, any pair of operations on $\mathbb{N}_{0}$ forming
a semiring on $\mathbb{N}_{0}$ can be used to define a semiring on multisets, by applying the operations to the multiplicity values in an element-wise manner.
\begin{definition}[Multiset Semiring]\label{def:multiset-semiring}
Given a semiring
\[\mathbb{K}_{\mathbb{N}_{0}} =(\mathbb{N}_{0},\oplus_{\mathbb{N}_{0}},\otimes_{\mathbb{N}_{0}},\overline{0}_{\mathbb{N}_{0}},\overline{1}_{\mathbb{N}_{0}})\, \text{on}\, \mathbb{N}_{0}\]
we define the \textit{multiset semiring} 
\[\mathbb{K}_{\mathcal{M}} =(\mathcal{M}_{A},\oplus_{\mathcal{M}_{A}},\otimes_{\mathcal{M}_{A}},\overline{0}_{\mathcal{M}_{A}},\overline{1}_{\mathcal{M}_{A}})\] on $\mathbb{K}_{\mathbb{N}_{0}}$ as:
\begin{itemize}
    \item ($M_1 \oplus_{\mathcal{M}_{A}} M_2)(k) = M_1(k) \oplus_{\mathbb{N}_{0}} M_2(k)
    \quad\forall k \in K$,
    \item ($M_1 \otimes_{\mathcal{M}_{A}} M_2)(k) = M_1(k) \otimes_{\mathbb{N}_{0}} M_2(k)
    \quad\forall k \in K$,
    \item $\overline{0}_{\mathcal{M}_{A}} = \{ a \rightarrow \overline{0}_{\mathbb{N}_{0}}\,|\,a \in A \}$, and
    \item $\overline{1}_{\mathcal{M}_{A}} = \{ a \rightarrow \overline{1}_{\mathbb{N}_{0}}\,|\,a \in A \}$.
\end{itemize}
\end{definition}
Finally, we define
$$M_1 \subseteq M_2\,\Leftrightarrow\,M_1(a) \leq M_2(a) \quad\forall a \in A$$
to lift the subset relation to multisets as usual.

\begin{figure}[pt]
    \centering
    \begin{tikzpicture}[->, >={Stealth[length=2mm]}, auto, node distance=3.5cm, semithick]
    \tikzstyle{every state}=[fill=none, draw=black, text=black]

    \node[state, initial]     (A)                     {$q_1$};
    \node[state]              (B) [right=2.5cm of A]  {$q_2$};
    \node[state]              (C) [below=1.3cm of B]  {$q_3$};
    \node[state]              (D) [left=2.5cm of C]   {$q_4$};
    \node[state, accepting]   (E) [left=2.2cm of D]   {$q_5$};

    \path (A) edge                node[above, align=left, yshift=-2pt]                  {addTeam\\\{$Team^1$\}}                          (B)
          (B) edge [loop right]   node[above, align=left, xshift=5pt, yshift=1pt]       {addPlayer\\\{$Player^1$\}}                      (B)
              edge                node[right, align=left]                               {addTeam\\\{$Team^1$\}}                          (C)
              edge [bend left]    node[left, align=right, xshift=13pt, yshift=17pt]     {addSolitaire\\\{$Player^1$, $Solitaire^1$\}}    (D)
          (C) edge [loop right]   node[below, align=left, xshift=5pt, yshift=-3pt]      {addPlayer\\\{$Player^1$\}}                      (C)
              edge                node[below, align=center, yshift=2pt]                 {addChess\\\{$Player^2$, $Chess^2$\}}            (D)
          (D) edge [bend left]    node[below, align=center, yshift=2pt]                 {addBT\\\{$ProcMod^1$, $BT^1$\}}                 (E)
              edge                node[above, align=center, yshift=-2pt]                {addWiFi\\\{$WiFi^3$\}}                          (E)
              edge [bend left]    node[left, align=right, yshift=7pt]                   {addProcMod\\\{$ProcMod^2$\}}                    (A);
\end{tikzpicture}%
    \caption{Weighted Automaton for the Multiplayer Game.}
    \label{fig:wa-example}
\end{figure}
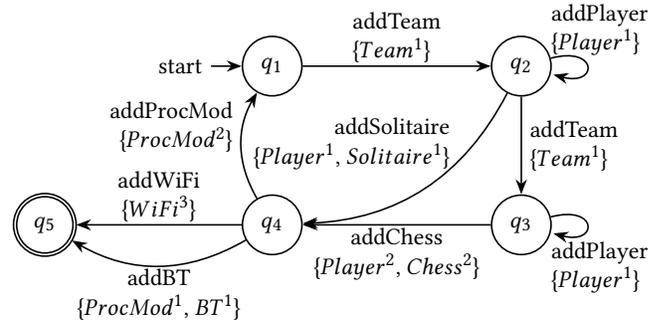

\smallskip
\noindent\textbf{Featured Multiset Semirings.}
Figure~\ref{fig:wa-example} shows an extract of a weighted automaton 
$\mathcal{A}_{\textit{MPG}}$ for the behavior of the Multiplayer Game
with the same structure and action alphabet 
as the FTS in Figure~\ref{fig:fts-example}.
Note that the shown extract actually describes the configuration
process of a game rather then the behavior of the game itself (see
Figure~\ref{fig:wa-example2} and Figure~\ref{fig:wa-decision} below).
However, we use this example as it contains most crucial cases in a graspable way.
Instead of Boolean presence conditions, behavioral variability is specified as transition weights over featured multisets $M\in\mathcal{M}_{F}$.
This representation is agnostic to the actual semiring used for 
mapping multisets over features to weighted automata.
\begin{table}[tp]
\centering
\caption{Overview of Numeric Semirings~\protect\cite{droste2009handbook}.}
\def\arraystretch{1.2}
\begin{tabular}{|l|c|c|c|c|}
\hline
Semiring     & $\oplus$ & $\otimes$ & $\overline{0}$ & $\overline{1}$ \\
\hline
Max-tropical & $max$    & $+$       & $-\infty$      & $0$            \\
Min-tropical & $min$    & $+$       & $\infty$       & $0$            \\
Min-min      & $min$    & $min$     & $\infty$       & $\infty$       \\
Max-max      & $max$    & $max$     & $-\infty$      & $-\infty$      \\
\hline
\end{tabular}
\label{tbl:semirings}
\end{table}
Table~\ref{tbl:semirings} shows a selection of
semirings~\cite{droste2009handbook} we consider as reasonable choices.

\smallskip
\noindent\textit{Max-Tropical Semiring.}
Using the \textit{max-tropical semiring} over
featured multisets, weights of transitions describe how many 
\textit{additional} feature instances are
required when taking the respective transition in a run
(i.e., every further occurrence of a transition in a path
increases the number of feature instances required 
in the respective configuration by the transition weight).
Note that we omit all features having weight $0$.
For instance, the weight $\{Team^{1}\}$ of the \textit{addTeam} transition denotes that the required number of instances of feature \textit{Team} increases by $1$ for every further traversal of this transition in a run. 
For each instance of feature \textit{Team}, exactly one \textit{Gamemode} must be selected thus requiring a corresponding number of instances of chess or solitaire games.
The \textit{addSolitaire} transition increments the number of instances of feature \textit{Solitaire} as well as of feature \textit{Player} to ensure that at least one player plays solitaire.
Similarly, Chess requires at least two players.
The game mode must be configured for each team separately according to the CFM, such that the number of instances of feature \textit{Chess} is also incremented by $2$ to cover both teams.
The \textit{addProcMod} transition adds two \textit{ProcMod} feature instances, as each additional game requires at least two more processing modules.
Moreover, the behavioral model specifies that three instances of feature \textit{WiFi} are required when choosing this communication method, whereas the number of processing modules is not restricted. 
The aggregated weight $M_{\ell}$ for a word $\ell$ corresponds to 
the overall sum of feature instances required for accepting word $\ell$.
In case multiple accepting paths exist for $\ell$, the element-wise \textit{max}-operation selects for each feature the maximum
number of instances required along all accepting paths.
This ensures that the final weight (multiset) reflects a sufficient number of feature instances to accept word $\ell$ in any case.
For instance, the aggregated weight for accepting the word
\begin{equation*}
    \begin{split}
\ell = &\,addTeam, addSolitaire, addProcMod, addTeam, addTeam,\\ 
       &\,addChess, addBT
\end{split}
\end{equation*}
in the \textit{max-tropical semiring} is
\[M_{\ell} = \{BT^1, ProcMod^3, Team^3, Player^3, Solitaire^1, Chess^2\}\]

\begin{figure}[pt]
    \centering
    \begin{tikzpicture}[->, >={Stealth[length=2mm]}, auto, node distance=3.5cm, semithick, initial text=start\\\{$Player^2$\}, align=center]
    \tikzstyle{every state}=[fill=none]

    \node[state, accepting]               (C)                   {$q_3$};
    \node[state]                          (D) [right=2cm of C]  {$q_4$};
    \node[state, accepting]               (E) [right=2cm of D]  {$q_5$};
    \node[state, initial, accepting]      (A) [above=1cm of C]  {$q_1$};
    \node[state, accepting]               (B) [above=1cm of E]  {$q_2$};

    \path (A) edge [bend left=20]    node[above, align=center] {movePlayer1}                      (B)
              edge                   node[right, align=center] {playBrainHand\\\{$Player^1$\}}    (C)
          (B) edge                   node[above, align=center] {movePlayer2}                      (A)
          (C) edge                   node[above, align=center] {brain}                            (D)
          (D) edge                   node[above, align=center] {hand}                             (E)
          (E) edge [bend left=20]    node[below, align=center] {movePlayer2}                      (C);
\end{tikzpicture}%
    \caption{Weighted Automaton for Variable Behaviour during a Chess Game.}
    \label{fig:wa-example2}
\end{figure}
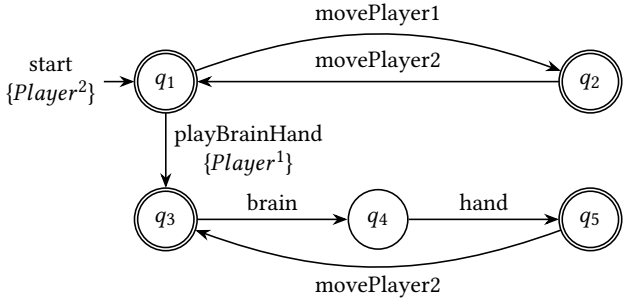
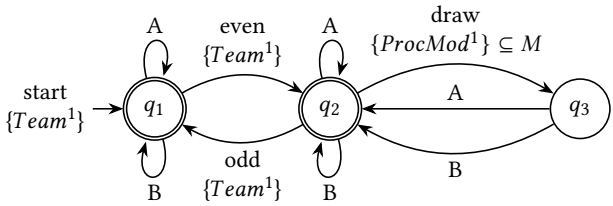
\begin{figure}[pt]
    \centering
    \begin{tikzpicture}[->, >={Stealth[length=2mm]}, auto, node distance=3.5cm, semithick, initial text=start\\\{$Team^1$\}, align=center]
    \tikzstyle{every state}=[fill=none]

    \node[state, initial, accepting]    (A)                      {$q_1$};
    \node[state, accepting]             (B) [right=1.5cm of A]   {$q_2$};
    \node[state]                        (C) [right=2.5cm of B]   {$q_3$};

    \path (A) edge [bend left]    node[above, align=center] {even\\\{$Team^1$\}}                  (B)
              edge [loop above]   node[above, align=center] {A}                                   (A)   
              edge [loop below]   node[below, align=center] {B}                                   (A)  
          (B) edge [bend left]    node[below, align=center] {odd\\\{$Team^1$\}}                   (A)
              edge [loop above]   node[above, align=center] {A}                                   (B)   
              edge [loop below]   node[below, align=center] {B}                                   (B)
              edge [bend left]    node[above, align=center] {draw\\$\{ProcMod^1\}\subseteq M$}    (C)
          (C) edge                node[above, align=center] {A}                                   (B)
              edge [bend left]    node[below, align=center] {B}                                   (B);
\end{tikzpicture}%
    \caption{Weighted Automaton for Variable Team Behaviour depending on the Number of Teams.}
    \label{fig:wa-decision}
\end{figure}

\smallskip
\noindent\textit{Max-Max Semiring.}
In the \textit{max-max} semiring, both semiring operations are defined as max-operation, such that we ignore all features having weight of $-\infty$.
Here, weights describe how many 
feature instances are \textit{at least}
required for executing the respective transition in a run.
The weight of a word along one accepting path 
corresponds to the \textit{maximum lower bound} 
for the number of required instances per feature.
The final weight for word $\ell$ aggregated from 
all accepting paths thus 
describes how many feature instances are \textit{at least} required
for all possible accepting paths of word $\ell$.
Hence, the weight for the word
\begin{equation*}
    \begin{split}
\ell = &\,addTeam, addSolitaire, addProcMod, addTeam, addTeam,\\ 
    &\,addChess, addBT
\end{split}
\end{equation*}
in the \textit{max-max semiring} is
\[M_{\ell} =\{BT^1, ProcMod^2,
Team^1, Player^2,
Solitaire^1, Chess^2\}\]

\smallskip
\noindent\textit{Min-X Semiring.}
In the \textit{min-tropical} semiring, the overall
number of feature instances is given as the minimum number aggregated over all accepting paths and in the \textit{min-min} semiring, the minimum operation is also used along one particular path, respectively, to obtain the \textit{minimum upper bound}.
In contrast to the \textit{max-tropical} and \textit{max-max} semirings in which
the weight of a word may be interpreted as \textit{costs} 
(i.e., we have to buy at least this number of feature instances
to enable this run), the \textit{Min-X} case may be interpreted as \textit{rewards} 
(we gain at most this number of feature instances by this run).
To this end, we may partition the set $F$ of features into
multiple subsets and apply separate semirings
to aggregate their costs.
For example, the weight of the word
\begin{equation*}
    \begin{split}
\ell = &\,addTeam, addSolitaire, addProcMod, addTeam, addTeam,\\ 
    &\,addChess, addBT
\end{split}
\end{equation*}
in the \textit{min-min semiring} is
\[M_{\ell} =\{BT^1, ProcMod^1,
Team^1, Player^1,
Solitaire^1, Chess^2\}\]

\smallskip
\noindent\textbf{Example: Hand and Brain.}
Figure~\ref{fig:wa-example2} shows an extract from the variable behavior during a chess game. 
Initial state $q_{1}$ requires as initial weight two players to start a chess game. 
The two players make alternating moves until one player wins and the game ends, which may happen after an arbitrary number of moves as $q_{1}$ and $q_{2}$ are both accepting states.
After any two consecutive moves, it is also possible that 
a third player joins the game to play hand and brain (i.e., one player plays against two players of which one appoints the chess piece to be played in a move and the other one thinks about how to move the piece).

\smallskip
\noindent\textbf{Composite Featured Multiset Semirings.}
To combine and simultaneously evaluate different interpretations of
weights for features into one behavioral model, multiple (independent) semirings 
can be composed into composite semirings~\cite{manger2008catalogue}.
For instance, we may combine the max-max and min-min semiring to track 
the maximum lower bound as well as minimum upper bound of instances for each feature.
If the minimum upper bound weight for any feature is less than the
maximum lower bound for that feature, then the word should be considered invalid.
To generalize, we consider an arbitrary collection 
$\mathbb{K}_{F_{1}}, \mathbb{K}_{F_{2}},\ldots, \mathbb{K}_{F_{n}}$ of $n$
different semirings $\mathbb{K}_{F_{i}}$, $1\leq i\leq n$, 
each defined over a subset $F_i \subseteq F$.
For the sake of plausibility, we require $F=\bigcup_{1\leq i\leq n}F_i$,
whereas we do not require the subsets to be mutually disjoint.
This generalization is handled by \textit{multi-weighted automata}~\cite{droste2016multi} 
in which weights are tuples from the Cartesian product of multiple independent weight components.
The weighted language is obtained by component-wise semiring operations.

\smallskip
\noindent\textit{Example: Encoding FTS.}
In the previous example, we used an X-tropical semiring
to aggregate increments of numbers of feature instances depending on the length of words. 
This allows us to express that the number of feature instances
required for accepting a word depends on the number of occurrences 
of particular actions, which is, in case of unbounded features, not expressible by FTS.
Conversely, the X-tropical multiset semiring is not able to express 
feature constraints on words expressed by path constraints 
aggregated from the sequence of presence conditions over Boolean features along that path of a run.
Multi-weighted automata over composite featured multiset semirings allow us
to combine different types of aggregation operators for different types 
of features (e.g., Boolean vs. multi-instance features).
We use a composite semiring combining the \textit{max-max} semiring and \textit{min-min} semiring 
both over the subset of Boolean features to encode Boolean presence conditions as composite weights.

\begin{figure}[pt]
    \centering
    \begin{tikzpicture}[->, >={Stealth[length=2mm]}, auto, node distance=3.5cm, semithick]
    \tikzstyle{every state}=[fill=none, draw=black, text=black]

    \node[state]              (D) []                    {$q_4$};
    \node[accepting, state]   (E) [right=5cm of D]      {$q_5$};

    \path (D) edge [bend right=20] node[below, align=center, yshift=1pt]   {addBT\\$\{ProcMod^1,\,BT^1\} \subseteq M \subseteq \{ProcMod^1,\,BT^1,\,WiFi^0\}$}    (E)
              edge                 node[above, align=center, yshift=-1pt]  {addWiFi\\$\{WiFi^3\} \subseteq M \subseteq \{WiFi^3,\,BT^0\}$}                        (E);
\end{tikzpicture}%
    \caption{Multi-Weighted Automaton for an Extract of the Multiplayer Game.}
    \label{fig:wa-example-minmax}
\end{figure}
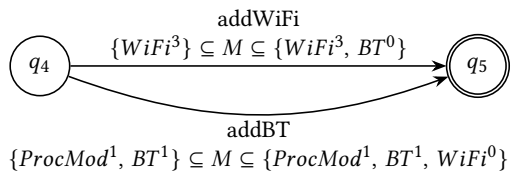

Figure~\ref{fig:wa-example-minmax} shows an extract of this construction for our example 
for the choice between \textit{Bluetooth} and \textit{WiFi}.
We denote both bounds in one line using 
sub-multiset operators to restrict multiset configuration $M$.
We omit unrestricted features (e.g., \textit{Player} 
has lower bound $-\infty$ and upper bound $\infty$).
As feature \textit{BT} has the upper bound $0$ for the \textit{addWiFi} transition, this transition excludes this feature, and the converse case holds for feature \textit{WiFi} for the \textit{addBT} transition, thus expressing logical exclusion between both features.
The lower- and upper-bound weights of the \textit{addWiFi} transition further specify that exactly $3$ additional instances of the unbounded feature \textit{WiFi} are required for every occurrence of this transition in a run.

\smallskip
\noindent\textbf{Example: Even-Odd.}
Figure~\ref{fig:wa-decision} shows another extract from the 
behavior of the multiplayer game. 
Initially, the number of teams is $1$ (odd) when starting in state $q_1$. During a game, the number of teams can be dynamically incremented by $1$ without upper bounds, such that the current number of teams alternates
between an even and an odd number. 
During a game, the teams frequently have to make collaborative choices between some alternatives $A$ and $B$, represented by transitions of state $q_1$ (odd number of teams) and state $q_2$ (even number of teams) states.
In case the number of teams is odd, the majority decision always wins, whereas in case of an even number, the decision may be inconclusive
due to a draw. 
In this case, a processing module is required to mediate the decision process.
In this example, we use the max-tropical semiring for feature \textit{Team} 
and the max-max semiring for feature \textit{ProcMod}.

\subsection{Mapping Configuration Spaces to Weighted Languages}\label{sec:main-mapping-approach}

We next discuss various aspects of defining mappings between
weighted words and multiset configurations.

\smallskip
\noindent\textbf{Configuration-Space Mapping.}
Let $F$ be a set of features, $\textit{cm}$ a CFM over $F$,
and $\mathcal{A}$ a weighted automaton over a multiset semiring $\mathbb{K}_{F}$ on $F$.
The weighted language accepted by 
$\mathcal{A}$ defines a mapping $\mathcal{A}:\Sigma^{*}\rightarrow K$ from words $\ell\in\Sigma^{*}$
to weights $k$ if $(\ell,k)\in\mathcal{L}(\mathcal{A})$, where $k$ 
is a (composite) multiset $M$ over $F$.
Similar to path conditions of FTS, 
multiset $M$ defines aggregated constraints to be satisfied by a configuration $M'\in\llbracket\textit{cm}\rrbracket$
to be able to perform an accepting run for $\ell$.
If $M$ aggregates the maximum lower bound per feature,
then we require $M' \supseteq M$ (i.e., $M'$ contains at least the number 
of instances per feature as required by $M$).
Conversely, if $M$ aggregates the minimum upper bound per feature,
then we require $M' \subseteq M$ (i.e., $M'$ contains at most the number 
of instances per feature as required by $M$).
In this way, we define a mapping of words to 
sets of multiset configurations accepting these words.
The opposite direction is defined in terms of \textit{projection}.

\smallskip
\noindent\textbf{Solution-Space Projection.}
By $\mathcal{A}|_{M}$, we denote the 
weighted automaton projected from $\mathcal{A}$
for multiset configuration $M\in\llbracket\textit{cm}\rrbracket$.
The weighted language of projection $\mathcal{A}|_{M}$
is restricted to those weighted words accepted by $M$.
If we use min-/max-operators for aggregating weights along paths
(e.g., min-min and max-max semiring), the projection 
can be constructed purely syntactically, by 
checking for each transition in $\mathcal{A}$ individually whether $M$ satisfies the weight constraint (and by removing the transition from $\mathcal{A}|_{M}$, otherwise).
This is similar to FTS projection~\cite{classen2012featured}.
In case of the X-tropical semiring, however, 
such a purely local construction of a projection 
$\mathcal{A}|_{M}$ is not feasible.
For example, using an X-tropical semiring, transition weights are added along paths.
Hence, it is not sufficient to check whether $M$ satisfies the weight constraint of each transition individually.
Instead, one and the same transition in $\mathcal{A}$ may be part
of a path for which $M$ satisfies the aggregated weight as well as
of another path for which $M$ does not satisfy the aggregated weight.
Hence, the construction of $\mathcal{A}|_{M}$ would require
to duplicate particular model parts of $\mathcal{A}$ 
such as shown for the FTS example in Figure~\ref{fig:fts-example-unrolled}.

\smallskip
\noindent\textbf{Solution-Space Analysis.}
When using FTS as behavioral variability model~\cite{classen2012featured}, 
there may exist paths in the transition graph having path conditions that contradict
the feature model (e.g., the path contains one transition
with presence condition $f$ and another with
presence conditions $f'$, where $f$ and $f'$ exclude
each other in the feature model).
Conversely, there may be valid configurations of the feature model
for which no valid path exists in the transition graph
(e.g., configurations in which some optional feature $f$ 
is not selected although it is required
on every non-empty path of the transition graph).
Similar cases may arise when using weighted 
automata over featured multisets as behavioral variability model.
Based on configuration-space mapping and solution-space projection,
we characterize \textit{mapping-consistency} in two directions:
\begin{itemize}
    \item Given a word $\ell$ with an accepting run in $\mathcal{A}$, does there exist a valid configuration $M$ satisfying the weight $k=\mathcal{A}(\ell)$?
    \item Given a valid configuration $M$, does there exist an accepting run in $\mathcal{A}$ such that $M$ satisfies the weight $k=\mathcal{A}(\ell)$?
\end{itemize}
An obvious advantage of weighted automata as target
mapping formalism is their rich body of theoretical knowledge
about canonical analysis problems and corresponding
complexity/decidability properties~\cite{almagor2022s}.
For instance, the second problem coincides with 
the \textit{non-emptiness problem} for weighted automata.
Further interesting decision problems include:
\begin{itemize}
    \item \textbf{Emptiness.} Does a given configuration $M$
    accept \textit{no} words accepted by $\mathcal{A}$?
    \item \textbf{Universality.} Does a given configuration $M$
    accept \textit{all} words accepted by $\mathcal{A}$?
    \item \textbf{Upper boundedness.} Does there exist a 
    configuration $M$ for which universality holds?
    \item \textbf{Lower boundedness.} Does there exist a 
    configuration $M$ for which emptiness holds?
\end{itemize}
Complexity/decidability of these problems
depend on the underlying semiring used and on
whether the automaton is deterministic or non-deterministic.
All semirings considered in this paper (see Table~\ref{tbl:semirings})
have feasible complexity in this regard, at least in the deterministic case.
This is mostly due to the discrete value domain $\mathbb{N}_{0}$ underlying 
multisets and the algebraic properties of all considered semirings.
For deterministic weighted automata over the value domain $\mathbb{N}_{0}$, the considered decision problems are in PTIME. For non-deterministic weighted automata over $\mathbb{N}_{0}$ (e.g., Example in Figure~\ref{fig:wa-simple-example}), the considered decision problems are PSPACE-complete
(see Table~\ref{tbl:wa-problem-complexity}). We refer to \citeauthor{almagor2022s} for details about the complexity of these problems \cite{almagor2022s}.
\citeauthor{droste2009weighted} have also introduced model checking of properties based on weighted MSO logic going beyond the basic decision problems described here \cite{droste2009weighted}.
\begin{table}[tp]
\caption{Complexity Classes of Decision Problems on Deterministic and Non-Deterministic Weighted Automata over $\mathbb{N}_{0}$~\protect\cite{almagor2022s}.}
\def\arraystretch{1.2}
\begin{tabular}{|l|l|l|}
\hline
Decision problem          & Determ. $\mathbb{N}_{0}$-WAs & Nondeterm. $\mathbb{N}_{0}$-WAs \\ \hline
(Non-)Emptiness   & PTIME                          & PSPACE-complete                   \\
Universality      & PTIME                          & PSPACE-complete                   \\
Lower boundedness & PTIME                          & PSPACE-complete                   \\
Upper boundedness & PTIME                          & PSPACE-complete                   \\ \hline

\end{tabular}
\label{tbl:wa-problem-complexity}
\end{table}
\section{Implementation and Evaluation}\label{sec:evaluation}

\subsection{Implementation}\label{sec:evaluation-impl}

Our tool is able to check weighted automata (WA) over featured multiset semirings for (non-)emptiness, universality and upper/lower boundedness properties.

\smallskip
\noindent\textbf{JAutomata Library.}
Our tool is built upon the \textit{JAutomata} library \cite{JAutomata}
for weighted finite state automata in Java.
The library includes various general purpose WA algorithms, which are
based on computing the shortest paths using a path-traversal algorithm~\cite{mohri2002semiring, droste2009handbook}.

The library has a generic API that allows to apply 
practically any semiring that can be defined in Java.
WAs are defined using a Java API to select the semiring and instantiate the states and transitions with weights for the respective semiring.
Implementations of Boolean, real, min-tropical and log semirings are included.

We modified the JAutomata library in two important ways.
Firstly, the element ordering used for the path-traversal algorithm can now be
specified explicitly instead of being derived from the semiring.
This allows us to support additional semirings like max-tropical,
which have not been usable until now due to termination problems of the path-traversal algorithm in case of loops.
This extension allows us, for instance, to consider
different orderings for individual features.

Secondly, a weight filter condition can be specified to determine when
the path-traversal algorithm will skip a state/transition or terminate early.
In the original version, the search only terminates after reaching a specified
maximum number of paths. Furthermore, transitions were only skipped if this does
not change the final weight.
This obstructs efficiency of property analysis,
as the path-traversal algorithm would continue exploring paths
even when the property has definitively been evaluated to true/false already.

\smallskip
\noindent\textbf{Support for Multiset Semirings.}
We provide an implementation of multiset semirings, which allows
mapping an arbitrary key type (features) to a value type from any other semiring.
Features can be represented as Java Enumerations or Strings.
Composite semirings are provided to combine arbitrary pairs of semirings,
which can be used to represent any multi-weighted semirings
as nested pair-wise composite semirings.
We also provide implementations of min/max tropical semirings
and min-min/max-max semirings both on \textit{int} and \textit{double} types.
Our tool supports checks for non-emptiness, universality, lower boundedness and upper boundedness properties
on multiset semirings defined on the max-tropical value semiring.

\smallskip
\noindent\textbf{Bound Parameter.}
The path-traversal algorithm of the JAutomata library
used in our tool first requires a semiring conversion into
a $k$-tropical semiring over lists of paths including their weight.
The $k$-tropical semiring represents lists of paths,
limited to at most $k$ paths.
If an operation would result in more than $k$ paths,
then only the first $k$ paths are considered.
This \textit{bound parameter} $k$ must be set when calling
the path-traversal algorithm.
The correctness of the results thus depends on the choice of
$k$ being sufficiently large to cover all relevant paths,
where larger values naturally increase the runtime.
This parameter allows us to investigate trade-offs between scalability and
precision of analyses in our evaluation.

\subsection{Research Questions}

\begin{itemize}
    \item[\textbf{(RQ1)}] \textbf{Efficiency.} How does parameter 
    $k$ influence the computational effort for evaluating emptiness, universality,
    lower boundedness and upper boundedness of WA 
    over featured multiset semirings?
    \item[\textbf{(RQ2)}] \textbf{Effectiveness.} How does parameter 
    $k$ influence the precision for evaluating emptiness, universality,
    lower boundedness and upper boundedness of WA 
    over featured multiset semirings?
\end{itemize}

\subsection{Methodology}

We perform separate analysis runs for checking for 
\textit{emptiness}, \textit{universality},
\textit{lower boundedness} and \textit{upper boundedness} (see Section~\ref{sec:main-mapping-approach}) 
for each subject system.
Concerning \textit{emptiness} and \textit{universality}, 
some multiset configuration $M$ must be given for 
which the respective property is evaluated.
To obtain a finite number of configurations, 
we selected a set of $17$ configurations
and manually derived the ground truth for these configurations
for the considered properties.
These $17$ configurations cover different combinations of feature assignments that were deemed interesting with respect to the WA.
The properties \textit{lower boundedness} and \textit{upper boundedness}, instead,
are global properties that require no further input.
Concerning bound parameter $k$, we evaluate 
each experiment for \textbf{RQ1} using 
sample values $k \in \{ 500, 1000, ..., 2500 \}$.
We do not consider $k < 500$ for \textbf{RQ1}, as the runtime for $k = 500$
is already too short (around $50$ms per run) to allow for 
a proper visualization of measurements.
For \textbf{RQ2}, we evaluate each experiment with 
more fine-grained sample values $k \in \{ 100, 200, ..., 1500 \}$
than for \textbf{RQ1} to investigate in more detail the influence
of parameter $k$ on the precision of analysis results.
We consider the same order for prioritization
and exploration of paths in all experiments.
We consider an ordering for all experiments in which the number of instances of feature \textit{Player} is minimized in the first stage, which means that the shortest paths considered first require as few players as necessary. In the next stages, the number of instances of features \textit{Team} and, subsequently, the number of instances of feature \textit{ProcMod} are minimized. The order for the remaining features is based on the generic multisubset relation.

\smallskip
\noindent\textbf{Test Data.}
We consider as test data the Multiplayer Game example
shown in Figure~\ref{fig:wa-example} in which we
use the max-tropical featured multiset semiring.
We also consider three different mutants of this WA in separate experiments.
To obtain these mutants, we removed the following transitions:
1) \textit{addWiFi}, 2) \textit{addWiFi} and \textit{addChess},
3) \textit{addWiFi}, \textit{addChess} and \textit{addProcMod}.
The resulting WAs are presumably less complex 
to analyze than the original WA due to their reduced size.
As these WAs are deterministic (i.e., every word corresponds to at most one path), the results for the min-tropical featured multiset semiring are identical as the max/min operations are never applied.
We further created WAs for the Multiplayer Game,
using max-max and min-min semirings to represent lower and upper bounds.
We thereby investigate the feasibility of composite semirings
and projection (see Section~\ref{sec:main-mapping-approach}).

\smallskip
\noindent\textbf{Data Collection.}
To evaluate effectiveness, each result reporting whether a property
is satisfied or not is recorded and compared to a manually obtained 
ground truth.
To evaluate efficiency, the runtime required for each analysis run is recorded.
Each experiment is repeated 10 times in a row 
to reduce the impact of variations in the measuring environment.
The 10 runs of each experiment are preceded by 3 unrecorded warm-up runs.
The overall approach contains no non-deterministic steps.
Hence, it is feasible to evaluate effectiveness without repetitions and warm-up phase.

\smallskip
\noindent\textbf{Measuring Environment.}
All experiments were performed on a Windows 10 system with
an Intel i7-5820K processor, OpenJDK 22.0.1 and a maximum of 2 GiB of Java Heap Space.
Our modified and extended version of the JAutomata library is 
included in our artifact \cite{mueller2024mappingartifact}.

\subsection{Results}

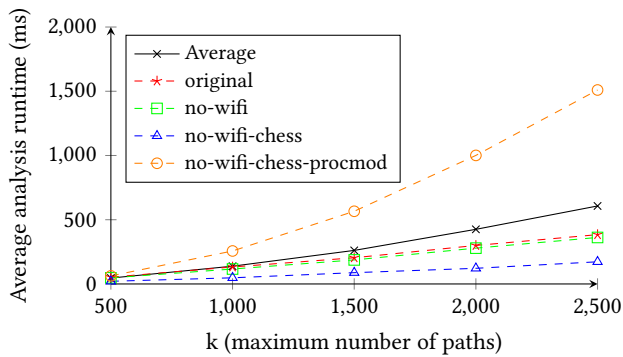
\begin{figure}[tp]
    \centering
    \begin{tikzpicture}
\begin{axis}[
	width=0.95\linewidth,
	height=5cm,
	legend pos=north west,
	legend cell align=left,
	legend style={nodes={scale=0.9, transform shape}},
	axis x line=bottom,
	axis y line=left,
	xtick={500, 1000, 1500, 2000, 2500},
	ytick={0, 500, 1000, 1500, 2000},
	xlabel={k (maximum number of paths)},
	ylabel={Average analysis runtime (ms)},
	ylabel style={above},
	xmin=500,
	xmax=2500,
	ymin=0,
	ymax=2000,
]

\addplot[black, mark=x] coordinates {
(500, 45.98472222)
(1000, 138.2666667)
(1500, 261.2097222)
(2000, 425.2888889)
(2500, 607.0555556)
};

\addplot[red, dashed, mark=star, mark options={solid}] coordinates {
(500, 52.90277778)
(1000, 132.1805556)
(1500, 204.0138889)
(2000, 299.7722222)
(2500, 383.8638889)
};

\addplot[green, dashed, mark=square, mark options={solid}] coordinates {
(500, 47.37222222)
(1000, 116.2777778)
(1500, 187.1194444)
(2000, 278.8583333)
(2500, 361.9694444)
};

\addplot[blue, dashed, mark=triangle, mark options={solid}] coordinates {
(500, 20.82222222)
(1000, 47.86388889)
(1500, 87.92777778)
(2000, 121.8333333)
(2500, 172.5111111)
};

\addplot[orange, dashed, mark=o, mark options={solid}] coordinates {
(500, 62.84166667)
(1000, 256.7444444)
(1500, 565.7777778)
(2000, 1000.691667)
(2500, 1509.877778)
};

\addlegendentryexpanded{Average}
\addlegendentryexpanded{original}
\addlegendentryexpanded{no-wifi}
\addlegendentryexpanded{no-wifi-chess}
\addlegendentryexpanded{no-wifi-chess-procmod}

\end{axis}
\end{tikzpicture}%
    \caption{Efficiency Results (RQ1) for the individual Weighted Automata and Total Average.}
    \label{fig:results-efficiency}
\end{figure}

\noindent\textbf{RQ1.}
The results concerning efficiency are shown in Figure~\ref{fig:results-efficiency}.
The x-axis denotes parameter $k$ and the y-axis denotes 
the average runtime in milliseconds 
for the analysis of all four properties.
One data series is shown for each variant of the WA (dashed, colored)
as well as for the average of all variants (solid, black).
With $k \leq 500$, the average runtime is below $50$ms.
The runtime for \textit{original}, \textit{no-wifi}
and \textit{no-wifi-chess} increases mostly linearly,
up to $384$ms, $362$ms and $173$ms, respectively, for $k=2500$.
The runtime for \textit{original} and \textit{no-wifi}
is almost equal, with runtime for \textit{no-wifi}
being around $10\%$ lower than \textit{original}
and runtime for \textit{no-wifi-chess} being around
$60\%$ lower than \textit{original} for all considered $k$.
The runtime for \textit{no-wifi-chess-procmod}
is on average around $163\%$ higher than for \textit{original}
and increases faster than for the other WAs, reaching 1510ms at $k = 2500$.
The runtime for \textit{no-wifi-chess-procmod}
increases by a factor of $2.4$ on average for each $500$-increment of $k$,
whereas the runtime for the other three WAs only increases by $1.7$ on average.

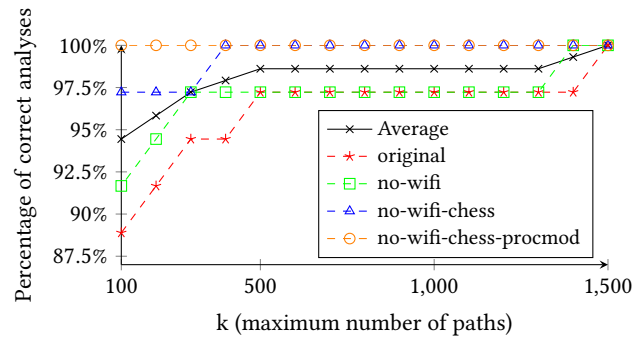
\begin{figure}[tp]
    \centering
    \begin{tikzpicture}
\begin{axis}[
	width=0.95\linewidth,
	height=4.5cm,
	legend pos=south east,
	legend cell align=left,
	legend style={nodes={scale=0.9, transform shape}},
	legend image post style={dash phase=0pt},
	axis x line=bottom,
	axis y line=left,
	xtick={100, 500, 1000, 1500},
	ytick={0.875, 0.90, ..., 1},
	yticklabel={\pgfmathparse{\tick*100}\pgfmathprintnumber{\pgfmathresult}\%},
	xlabel={k (maximum number of paths)},
	ylabel={Percentage of correct analyses},
	ylabel style={above},
	xmin=100,
	xmax=1500,
	ymin=0.87,
	ymax=1,
]

\addplot[black, mark=x] coordinates {
(100, 0.944444444)
(200, 0.958333333)
(300, 0.972222222)
(400, 0.979166667)
(500, 0.986111111)
(600, 0.986111111)
(700, 0.986111111)
(800, 0.986111111)
(900, 0.986111111)
(1000, 0.986111111)
(1100, 0.986111111)
(1200, 0.986111111)
(1300, 0.986111111)
(1400, 0.993055556)
(1500, 1)
};

\addplot[red, dashed, mark=star, mark options={solid}] coordinates {
(100, 0.888888889)
(200, 0.916666667)
(300, 0.944444444)
(400, 0.944444444)
(500, 0.972222222)
(600, 0.972222222)
(700, 0.972222222)
(800, 0.972222222)
(900, 0.972222222)
(1000, 0.972222222)
(1100, 0.972222222)
(1200, 0.972222222)
(1300, 0.972222222)
(1400, 0.972222222)
(1500, 1)
};

\addplot[green, dashed, mark=square, dash phase=9pt, mark options={solid}] coordinates {
(100, 0.916666667)
(200, 0.944444444)
(300, 0.972222222)
(400, 0.972222222)
(500, 0.972222222)
(600, 0.972222222)
(700, 0.972222222)
(800, 0.972222222)
(900, 0.972222222)
(1000, 0.972222222)
(1100, 0.972222222)
(1200, 0.972222222)
(1300, 0.972222222)
(1400, 1)
(1500, 1)
};

\addplot[blue, dashed, mark=triangle, mark options={solid}] coordinates {
(100, 0.972222222)
(200, 0.972222222)
(300, 0.972222222)
(400, 1)
(500, 1)
(600, 1)
(700, 1)
(800, 1)
(900, 1)
(1000, 1)
(1100, 1)
(1200, 1)
(1300, 1)
(1400, 1)
(1500, 1)
};

\addplot[orange, mark=o, dashed, dash phase=9pt, mark options={solid}] coordinates {
(100, 1)
(200, 1)
(300, 1)
(400, 1)
(500, 1)
(600, 1)
(700, 1)
(800, 1)
(900, 1)
(1000, 1)
(1100, 1)
(1200, 1)
(1300, 1)
(1400, 1)
(1500, 1)
};

\addlegendentryexpanded{Average}
\addlegendentryexpanded{original}
\addlegendentryexpanded{no-wifi}
\addlegendentryexpanded{no-wifi-chess}
\addlegendentryexpanded{no-wifi-chess-procmod}

\end{axis}
\end{tikzpicture}%
    \caption{Effectiveness Results (RQ2) for the individual Weighted Automata and Total Average.}
    \label{fig:results-effectiveness}
\end{figure}

\smallskip
\noindent\textbf{RQ2.}
The effectiveness results are shown in Figure~\ref{fig:results-effectiveness}.
The x-axis denotes parameter $k$ and the 
y-axis denotes the percentage of correctly evaluated properties.
One data series is shown for each variant of the WA (dashed, colored)
and the average of all variants (solid, black).
The percentage of correct properties increases for all WAs
with increasing $k$ values until it reaches $100\%$.
All considered properties are evaluated correctly with $k \geq 1500$,
hence larger values of $k$ are not considered for this research question.
The correctness for the WAs \textit{original}, \textit{no-wifi},
\textit{no-wifi-chess} and \textit{no-wifi-chess-procmod}
with $k = 100$ starts at $88.89\%$, $91.67\%$, $97.22\%$ and $100\%$, respectively.
Hence, the properties are evaluated correctly with all considered values of $k$
for \textit{no-wifi-chess-procmod}.
The properties on \textit{no-wifi-chess} are evaluated
correctly with $k \geq 400$.
The properties on \textit{no-wifi} are evaluated
correctly with $k \geq 1400$.
The properties on \textit{original} are evaluated
correctly only with $k \geq 1500$.
Values of $k \le 100$ were neglected to improve readability.

\subsection{Discussion}

\noindent\textbf{RQ1.}
The \textit{no-wifi} mutant only differs from \textit{original}
by the missing \textit{addWiFi} transition.
Thereby, removing the Boolean choice between Bluetooth and WiFi
decreases runtime of all analyses.
But, this effect is small as this transition can only be activated
at most once in \textit{original}.
In \textit{no-wifi-chess}, we also removed the
\textit{addChess} transition and thereby the choice between the two modes.
This causes a significant reduction of the average runtime 
compared to \textit{original}.
In \textit{no-wifi-chess-procmod}, we further removed 
the \textit{addProcMod} transition such that 
the only remaining loop is the \textit{addPlayer} transition.
However, the average runtime for this behavioral model is significantly
higher than for the others, which leads 
us to the conclusion that there is no obvious
relationship between size of the behavioral model and runtime of analysis.
Another possible influence on runtime is whether 
a particular property is actually satisfied or not.
As our implementation quickly terminates 
if a (monotone) property cannot be satisfied anymore,
those properties requiring an exhaustive search of all possible paths
presumably take longer.
Upper boundedness is only satisfied for
\textit{no-wifi-chess-procmod} which explains why analyzing this property
takes longer than for the other behavioral models.

We also have to take into account 
how many of the $17$ sample configurations
used for non-emptiness and universality are satisfied
for the individual behavioral models.
Universality is satisfied $9$ times for \textit{no-wifi-chess-procmod},
but only 5 times for \textit{no-wifi-chess}.
Evaluating universality takes longer if it is satisfied, 
as it requires checking all paths, whereas encountering one
counter-example for universality will immediately terminate the search.
However, this does not fully explain the increase in runtime
also observed for non-emptiness checking,
which is satisfied $13$ times for \textit{original} and
$11$ times for the other behavioral models.
The increase in average runtime for all properties
with \textit{no-wifi-chess-procmod}
is mostly caused by the ordering in which path exploration
prioritizes transitions.
We defined the ordering for all experiments such that the number of \textit{Player}
features is minimized in the first stage,
which means that the shortest paths considered first require
as few players as necessary.
However, the only way to traverse different paths
in \textit{no-wifi-chess-procmod} is to add more players
via the \textit{addPlayer} loop.
Only $3$ out of $17$ configurations used for evaluating non-emptiness
and universality restrict the number of players such that 
evaluation of most properties took longer due to the
additional irrelevant \textit{addPlayer} paths.

\smallskip
\noindent\textbf{RQ2.}
Increasing the value of $k$ leads to an increase in the precision of the results,
as correct evaluation requires traversals of specific paths.
However, setting $k$ practically to $\infty$ significantly increases runtime
due to the combinatorial explosion of the number of paths.
Results for \textit{original} have the lowest average correctness,
followed by \textit{no-wifi} and \textit{no-wifi-chess}.
Finally, results for \textit{no-wifi-chess-procmod} are correct for all values of $k$.
Hence, as expected, average correctness as well as initial correctness 
decrease with increasing behavioral model size.
The results for \textit{original} and \textit{no-wifi} have similar
correctness values, as \textit{no-wifi}
requires less path traversals due to the missing final choice
between Bluetooth and WiFi.
Correctness for \textit{no-wifi-chess} is
higher than for \textit{no-wifi} and $100\%$ can be reached
with a much lower value of $k$.
This is due to the missing \textit{addChess} transition, which halves the number
of paths in every loop iteration,
whereas removing \textit{addWiFi} only halves the number of total paths once.

Lastly, correctness for \textit{no-wifi-chess-procmod} is
$100\%$ for all values of $k$, as the \textit{addProcMod}
loop is missing such that only Solitaire can be selected with
an arbitrary number of players.
Out of the $17$ configurations considered for universality,
only $3$ restrict the number of players. 
These were only evaluated correctly with higher $k$ values than for the other behavioral models,
which is due to the prioritization that minimizes the number of \textit{Player} features.
For \textit{no-wifi-chess-procmod}, the player-related
universality configurations are evaluated correctly already with $k=100$,
as these sample configurations do not restrict
the number of players above $100$.

\subsection{Threats to Validity}

\noindent\textbf{Internal Validity.}
Due to the novelty of the approach, the primary goal of the performed evaluation was to show feasibility and to demonstrate the usability of existing tools. As such, the experimental scope is currently limited to a few parameters and subject systems. Furthermore, model-theoretic restrictions of the approach are left as an open question for future work.
Correctness of evaluation results depend on the
bound parameter value $k$ being sufficiently large such that all relevant paths are covered.
The results for RQ2 show how correctness of
some properties changes depending on $k$.
This limitation stems from the utilization of
the existing path-traversal algorithm from JAutomata.
Our future goal is to implement property evaluation without 
explicit bound parameter $k$.
The need to explicitly specify an order in which paths are explored is also a potential threat.
This choice may also influence runtime (RQ1) and precision (RQ2).
We expect the order during path exploration to influence the results which we plan to further investigate as a future work.
Moreover, as the considered subject systems are all deterministic WAs, 
we did not encounter or evaluate the problem-intrinsic scalability issues due to non-determinism.
In the examples considered so far, we have not yet seen any need to express nondeterministic behavior. We argue that restricting the modeling to only deterministic WAs might be expressive enough, so the scalability issues might be neglectable.

\smallskip
\noindent\textbf{External Validity.}
We only investigated a small number of self-created 
subject systems in our evaluation due to the novelty of the proposed behavioral model,
including four different variants of the Multiplayer Game.
WAs over featured multisets are currently described directly in Java code which of course obstructs usability and adoption of the approach.
We plan to develop a text-based description format as a future work.
Moreover, we limited the number of manually chosen configurations for evaluating
the emptiness and universality properties to 17,
to ensure a reasonable limit for the overall 
preparation and execution time of the evaluation.
The selection of 17 configurations is therefore also a potential threat. As previously discussed for RQ1, whether or not the non-emptiness and universality properties are satisfied for specific configurations influences the runtime of the approach, hence a biased selection of configurations may lead to skewed results.
Finally, we did not compare our approach to any other approach. 
The most closely related approaches, 
FTS and FWA (see Section~\ref{sec:relatedwork}),
are, however, conceptually too different to our approach to allow for any meaningful comparison.
\section{Related work}\label{sec:relatedwork}

To the best of our knowledge, no approaches for mapping CFMs to 
behavioral variability modeling in the solution space have been proposed so far. 
In fact, there is even neither a commonly agreed upon syntax nor semantics for CFMs yet. 
We split our overview of related work into 
problem-space modeling formalisms including notions of multiplicity
and solution-space variability modeling formalisms in general.

\smallskip
\noindent\textbf{Variability Modelling with Cardinalities in the Problem Space.}
Several works use multiset based configuration semantics for CFMs, as we do. 
For instance, \citeauthor{weckesser2016mind} show that multiset based configuration semantics are sufficient to automatically 
detect anomalies such as interval gaps and false bounds of intervals in CFMs~\cite{weckesser2016mind}. 
\citeauthor{sampling2024} propose sampling criteria for CFM also based on a multiset based configuration semantics \cite{sampling2024}. 

Semantic considerations about CFMs are mostly concerned with how to
interpret cross-tree constraints. 
Using multiset based configuration semantics, a global interpretation is applied which takes 
the overall number of all feature instances into account~\cite{weckesser2016mind}.
Alternative interpretations use a local scope
(i.e., considering the number of instances per cloned subtree) and so-called relative constraints 
in the literature \cite{sousa2016extending}.
However, the solution-space mapping proposed in the paper is orthogonal to
these considerations.

Tool support for modeling and analyzing CFMs is sparse. 
Autonomous anomaly detection is implemented in the tool CardyGAn~\cite{schnabel2016cardygan}. 
This tool also provides modeling capabilities for CFMs and generation of valid configurations
as satisfiability witnesses. 
Clafer is another tool covering CFM modeling and analysis~\cite{juodisius2018clafer, bkak2016clafer}
which unifies feature modelling with class models and supports variability modelling with cardinalities.
However, both projects appear to be discontinued and none of them
incorporate solution-space variability mapping for CFMs.

\smallskip
\noindent\textbf{Behavioral Variability in the Solution Space.}
Annotation-based behavioral variability modeling formalisms in the solution space have been extensively studied and surveyed, but mostly for Boolean features only~\cite{benduhn2015survey, ter2019quantitative}. 
As the most generic approach, featured transition systems (FTS)~\cite{classen2012featured}
annotate transitions with Boolean presence conditions to express behavioral variability.
Featured Finite State Machines are very similar to FTS (i.e., also using Boolean presence conditions), yet providing richer syntactic constructs for behavioral modeling~\cite{hafemann2016validated}.
Another automata-based variability modeling approach are modal transition systems (MTS) \cite{fischbein2006foundation}, which distinguish between may- and must-transitions, 
thus reflecting mandatory and optional problem-space variability. 
Thereupon, many variations of modal transition systems have been proposed. 
Coherent modal transition systems with constraints \cite{ter2016modelling} add constraints 
to make MTS equally expressive as FTS. 
Modal Interface Automata have been proposed for family-based conformance testing of product families \cite{luthmann2015towards} and Featured Modal Contract Automata \cite{basile2017orchestration} unify Service-Oriented Computing with variability. 
Parametric Modal Transition Systems \cite{benevs2011parametric} enable the expression of exclusive, conditional and persistent choices to ensure compliance with previously made variability decisions.
All these MTS-based approaches are at most as expressive as FTS and thus not applicable as solution-space formalism for CFMs as discussed in Section~\ref{sec:background}.
Featured variants of timed automata (TA) have been proposed, called featured timed automata \cite{cordy2012behavioural} (FTA) and featured team automata \cite{ter2021featured}. 
Although the semantic domain of timed automata is inherently unbounded, both approaches 
are not able to express unbounded multiplicity and aggregated feature instances
as provided by our approach. In Feature Nets, a Petri net variant \cite{muschevici2016feature}, transitions are annotated by so-called application conditions denoting for which configuration this transitions is enabled.
Similar to FTS, application conditions are limited finite Boolean
configurations.

The most closely related formalism to our approach are Featured Weighted Automata (FWA)
\cite{fahrenberg2017featured, fahrenberg2019quantitative} which combine
FTS and WA into one model. 
Thus, FWA are more expressive than FTS by extending configuration-specific words 
by weights, but the underlying variability and mapping is still purely Boolean.
A naive extension of FWA by cardinality annotations for interfacing with CFMs, however,
would still be restricted to presence/absence of single transitions which is not sufficient to
handle unbounded cardinalities and aggregated feature instances (see Section~\ref{sec:background}).
\section{Conclusion}\label{sec:conclusion}

We presented a novel formalism for behavioral variability modeling
using weighted automata over featured multiset semirings
to capture infinite and non-convex configuration spaces shaped by CFMs.
Our experimental results show that 
the proposed behavioral model allows us to effectively evaluate essential semantic properties
for those behavioral models, at least for small subject systems.
Our tool implementation includes a bound parameter
which allows adjustments to also scale to larger-scale behavioral models
under reduced precision.

As a future work, we plan to extend and improve 
the formalism and the corresponding tool in various ways.
We also plan to describe best practices, modeling patterns and workflows to
support developers in manually creating models
using the proposed approach.
In addition, we aim for an approach for solution-space projection
on multiset configurations that also supports monotone operators
(e.g., max-tropical semiring).
Second, we plan to refine our tool to implement
weighted automata analyses without explicit bound parameters.
Third, we plan to use our formalism to 
reason about multi-objective NFP optimization of multiset
configurations using composite
semirings in multi-weighted automata.
This allows us to define novel sampling criteria
for CFMs based on solution-space knowledge~\cite{sampling2024}.
Finally, we wish to enlarge our collection of case studies 
to increase the corpus of weighted automata
for future evaluation.

\begin{acks}
We thank our reviewers for their constructive feedback. This work has been funded by the German Research Foundation within the project \textit{Co-InCyTe} (LO 2198/4-1).
\end{acks}

\bibliographystyle{ACM-Reference-Format}
\bibliography{paper}

\end{document}